\documentclass[preprint,aps,nofootinbib]{revtex4}

\usepackage{graphicx}

\begin{document}

\title{Constraint on $B-L$ cosmic string from leptogenesis with degenerate neutrinos}
\author{Pei-Hong Gu}
\email{guph@mail.ihep.ac.cn}
\author{Hong Mao}
\email{maohong@mail.ihep.ac.cn} \affiliation{Institute of High
Energy Physics, Chinese Academy of Sciences, P.O. Box 918-4,
Beijing 100049, People's Republic of China}
\begin{abstract}
In the early Universe, as a consequence of $U(1)_{B-L}$ gauge
symmetry-breaking, the so-called $B-L$ cosmic strings are expected
to be produced at the breaking scale $\eta _{B-L}$ according to
the Kibble mechanism. The decaying, collapsing closed loops of
these strings can release the right-handed neutrinos, whose
subsequent decay can contribute to the baryon asymmetry of the
Universe (BAU), through the "slow death"(SD) process and/or the
"quick death"(QD) process. In this paper, we assume that the decay
of the lightest heavy Majorana neutrinos released from the $B-L$
cosmic string loops can produce a baryon asymmetry consistent with
the cosmic microwave background (CMB) observations. Considering
the fact that both the neutrinoless double beta decay experiment
and the cosmological data show a preference for degenerate
neutrinos, we give the lower limits for the breaking scale
$\eta_{B-L}$ with the neutrino masses $0.06 \textrm{eV} \leq
\bar{m}=(m_{1}^{2}+m_{2}^{2}+m_{3}^{2})^{1/2} \leq 1.0
\textrm{eV}$, where the full possible cases of degenerate
neutrinos are included. We obtain $\eta_{B-L}\gtrsim 3.3 \times
10^{15} \textrm{GeV}$, $5.3 \times 10^{15} \textrm{GeV}$ and $9.5
\times 10^{15} \textrm{GeV}$ for $\bar{m}=0.2 \textrm{eV}$, $0.4
\textrm{eV}$ and $1.0 \textrm{eV}$ respectively in the SD process,
and find the $B-L$ cosmic string has a very small contribution to
the BAU in the QD process.
\end{abstract}
\maketitle

The baryon asymmetry of the Universe (BAU) has been determined
precisely\cite{tegmark}:
\begin{equation}
\eta _{B}^{CMB} \equiv \frac{n_{B}}{n_{ \gamma }}=(6.3 \pm 0.3)
\times 10^{-10},
\end{equation}
where $n_{B}= n_{b}-n_{ \overline{b}}$ and $n_{ \gamma }$ are the
baryon and photon number densities, respectively. At the same
time, there are several neutrino oscillation
experiments\cite{ahn,ahmad} which have confirmed the extremely
small but non-zero neutrino masses. Then
leptogenesis\cite{fukugita} is now an attractive scenario which
can simultaneously explain the cosmological baryon asymmetry and
the neutrino properties by the seesaw mechanism\cite{seesaw}.

The simplest leptogenesis scenario is to extend the standard model
(SM) by three generations of the right-handed neutrinos with
Majorana mass. A more appealing alternative is to consider this
within the context of unified models with an embedded $U(1)_{B-L}$
gauge symmetry, which can be derived from the $SO(10)$ models.
After spontaneous breaking of the $U(1)_{B-L}$ gauge symmetry, the
right-handed neutrinos naturally acquire heavy Majorana mass and
produce a lepton asymmetry, which finally converted to the baryon
asymmetry via the $(B-L)$-conserving sphaleron
process\cite{sphaleron}, by decaying into massless leptons and
electroweak Higgs bosons.

In the early Universe, according to the Kibble
mechanism\cite{kibble}, the so-called $B-L$ cosmic strings are
expected to be produced at the $U(1)_{B-L}$ symmetry-breaking
scale during $SO(10)$ breaking to the SM gauge
group\cite{kibble2}. The strings are formed by the gauge field and
Higgs field, and the Higgs field also gives heavy Majorana mass to
the right-handed neutrinos through Yukawa
coupling\cite{jackiw,weinberg}. As discussed in
Ref.\cite{bhattacharjee2,brandenberger,lew,jeannerot,bhattacharjee,sahu},
the decaying, collapsing closed loops of these strings can be a
nonthermal source of the right-handed neutrinos whose subsequent
decay can contribute to the BAU.

When the $B-L$ cosmic string loops contribute significantly to the
BAU, ones find that the $U(1)_{B-L}$ gauge symmetry-breaking scale
$\eta _{B-L}$ has a lower limit $\eta_{B-L} \gtrsim 1.7 \times
10^{11} \textrm{GeV}$\cite{sahu} if the light neutrino masses are
hierarchical, especially for $m_{3} \simeq 0.05 \textrm{eV}$,
where $m_{i}(i=1,2,3)$ is the eigenvalue of the light neutrino
mass matric. But once the evidence of neutrinoless double beta
decay with $m_{ee}=(0.05-0.86) \textrm{eV}$(at 95 \%
C.L.)\cite{dietz} is confirmed, a degenerate neutrino spectrum is
required. Recent studies on the cosmological
data\cite{spergel,tegmark} with the neutrino oscillation
experiment results\cite{ahn,ahmad} also showed a preference for
degenerate neutrinos with $m_{i}\lesssim 0.23 \textrm{eV}$ or
$m_{i}\lesssim 0.56 \textrm{eV}$.

In this paper, we follow the discussions in Refs.\cite{sahu} and
assume that the lightest heavy Majorana neutrinos are released
from the $B-L$ cosmic string loops, and hence a baryon asymmetry
consistent with the observations can be induced by their decay
modes. We estimate the $U(1)_{B-L}$ symmetry-breaking scale $\eta
_{B-L}$ for $0.06 \textrm{eV} \leq \bar{m} \leq 1.0 \textrm{eV}$,
where $\bar{m}$ is defined as
$\bar{m}^{2}=m_{1}^{2}+m_{2}^{2}+m_{3}^{2}$. In this range of
$\bar{m}$, the full possible cases of degenerate neutrinos are
included.

In the generic picture, local $B-L$ cosmic strings form at the
phase transition associated with the spontaneous symmetry breaking
of $U(1)_{B-L}$. During the phase transition, a network of strings
forms, consisting of both infinite strings and cosmic loops. After
the transition, the infinite string network coarsens and more
loops form from the intercommuting of infinite strings. In the
following discussion, we pay particular attention to the formation
of closed loops and their subsequent evolution.

The formation and evolution of the cosmic string loops have been
studied extensively, both in analytical and numerical methods, for
details see \cite{vilenkin,Hindmarsh}. After their formation, the
evolution of the closed loops can be broadly categorized into two
classes. One is the "slow death"(SD) process
\cite{jeannerot,vilenkin,Hindmarsh}, where the loops born at time
$t$ have a longer lifetime (compared to the Hubble expansion time
scale $H^{-1}(t)$). In this scenario, during the phase transition,
the formed string loops oscillate freely and loose their energy by
emitting gravitational radiation. When the loop's radius becomes of
the order of the string width, the loop releases its final energy
into massive particles. Among these particles will be the massive
gauge bosons, Higgs bosons and massive right-handed neutrinos which
were trapped in the string as fermion zero modes\cite{vilenkin}. The
other is the "quick death"(QD) process\cite{vincent}$-$the loops die
instantaneously as soon as they are formed due to the high
probability of self intersecting\cite{Siemens:1994ir}. Thus they
would lose only a negligible amount of energy in gravitational
radiation and massive particle radiation rather than gravitational
radiation plays the dominant role.

For the above two cases, the number density of loops disappearing in
the radiation dominated epoch at any time $t$ can be described
respectively as\cite{vilenkin,Hindmarsh,sahu}:
\begin{equation}
\frac{dn_{SD}}{dt}=f_{SD}\frac{1}{x^{2}}(\Gamma G
\mu)^{-1}\frac{(C+1)^{3/2}}{C}t^{-4},
\end{equation}
\begin{equation}
\frac{dn_{QD}}{dt}=f_{QD}\frac{1}{x^{2}}\mu ^{1/2}t^{-3},
\end{equation}
where $x$ is approximately in the range $\sim 0.4 - 0.7$ supported
by the extensive numerical simulations, $\Gamma \sim 100$ is a
geometrical factor that determines the average loop length, $ \mu
$ is the mass per unit length of a cosmic string and related the
symmetry-breaking scale $\eta$ with $\mu ^{1/2} \sim \eta$, $C$ is
a numerical factor of order unity and $G=1/M_{Pl}^{2}$ is the
Newton's constant, while $f_{SD}$ and $f_{QD}$ denote the fraction
of newly born loops which die through the SD process and QD
process respectively.

Noteworthy that the observations of the cosmic microwave
background (CMB) anisotropy give an upper bound on the
symmetry-breaking scale\cite{landriau}
\begin{equation}
\eta \lesssim 1.0 \times 10^{16} GeV.
\end{equation}
In addition, the measured flux of the cosmic gamma ray background
in the $10 \textrm{MeV} - 100 \textrm{GeV}$\cite{sreekumar} energy
region puts a constraint on $f_{QD}$\cite{Bhattacharjee:1998qc}
\begin{equation}
 f_{QD}(\eta /10^{16}GeV)^{2} \lesssim 9.6 \times 10^{-6},
\end{equation}
but there is no equivalent constraint on $f_{SD}$. This difference
can be understood easily, since the time dependence of the
disappearing rate of loops is $\propto t^{-4}$ in the SD case [see
Eq.(2)] while $\propto t^{-3}$ in the QD case [see Eq.(3)], in
other words, the SD process dominates at sufficiently early time,
while the QD process dominates at relatively late time and can
potentially contribute to the nonthermal gamma ray background.

It is difficult to calculate exactly the total number of the heavy
Majorana neutrinos from each loop, but as shown in
Ref.\cite{jeannerot} when a cosmic loop decays, it releases at least
one heavy Majorana neutrino. For simplicity, we may expect that it
would be a number of order unity. Then the releasing rate of the
heavy Majorana neutrinos $N _{i}(i=1,2,3)$ from the SD process and
QD process at any time $t$ can be written as:
\begin{equation}
\frac{dn_{N_{i}}^{SD}}{dt}=N_{N_{i}}^{SD}f_{SD}\frac{1}{x^{2}}(\Gamma
G \mu)^{-1}\frac{(C+1)^{3/2}}{C}t^{-4},
\end{equation}
\begin{equation}
\frac{dn_{N_{i}}^{QD}}{dt}=N_{N_{i}}^{QD}f_{QD}\frac{1}{x^{2}}\mu
^{1/2}t^{-3},
\end{equation}
with $N_{N _{i}}^{SD}$ and $N_{N_{i}}^{QD}$ $\sim O(1)$.

In the leptogenesis scenario, the decay of the heavy Majorana
neutrinos $N _{i}$ which produces the lepton asymmetry is described
by the following Lagrangian
\begin{equation}
-\mathcal{L}=h_{ij} \overline{l}_{Li} \phi \nu_{Rj} +
\frac{1}{2}M_{i} \overline{ \nu }_{Ri}^{c} \nu _{Ri} + H.c.,
\end{equation}
where $l$, $\phi $ are the SM leptons and Higgs doublets
respectively. In this framework, the heavy Majorana neutrino is
given by $N _{i}= \nu _{Ri} + \nu _{Ri}^{c}$ with heavy mass
$M_{i}$. The light neutrino mass matrix can be written as
\begin{equation}
m_{\nu}=-h^{\ast}\frac{1}{M}h^{\dag}v^{2}
\end{equation}
with $M=$diag$(M_{1},M_{2},M_{3})$ and $v= \langle \phi \rangle
\simeq 174\textrm{GeV}$.

For a hierarchical spectrum of the heavy Majorana neutrinos $M_{1}
\ll M_{2},M_{3}$, the lepton asymmetry, which is finally converted
to the baryon asymmetry, comes mainly from the decay of the lightest
heavy Majorana neutrino $N_{1}$ and the contribution of the cosmic
string loops to the BAU can be estimated as:
\begin{equation}
\eta _{B}=\frac{28}{79} \times 7.04 \times \varepsilon_{1} \int
^{t_{EW}} _{t_{F}} \frac{1}{s} (\frac{dn_{N_{1}}}{dt})dt,
\end{equation}
where $28/79$ is the value of $B/(B-L)$ for SM\cite{turner}, $7.04$
is the present density ratio of photon number and entropy,
$\varepsilon_{1}$ is the CP asymmetry of $N_{1}$ decays and $s=(2
\pi ^{2}/45)g_{\ast}T^{3}$ is the entropy density with $g_{\ast}
\simeq 106.75$. $t_{EW}$ is the electroweak transition time, while
$t_{F}$ denotes the epoch when the inverse decays and
$L$-nonconserving scatterings begin to freeze out and there is no
washout effects any more. $\frac{dn_{N_{1}}}{dt}$ is the releasing
rate of $N_{1}$ from the cosmic string loops. The temperature $T$
and time $t$ are related by
\begin{equation}
t=\frac{1}{2H(T)}
\end{equation}
with Hubble constant
\begin{equation}
H(T)=\sqrt{\frac{8 \pi^{3}g_{\ast}}{90}}T^{2}/M_{Pl}.
\end{equation}
Therefore we get
\begin{equation}
dt=-\frac{1}{TH(T)}dT
\end{equation}
and can rewrite Eq.(10) as
\begin{equation}
\eta _{B}=\frac{28}{79} \times 7.04 \times \varepsilon_{1} \int
^{T_{F}} _{T_{EW}} \frac{1}{s}
[\frac{dn_{N_{1}}}{dt}(T)]\frac{1}{TH(T)}dT.
\end{equation}

Now the key to calculate the baryon asymmetry in Eq.(14) is how to
determine $\varepsilon_{1}$ and $T_{F}$. We note that there is an
upper bound on $\varepsilon_{1}$ with
$m_{1}<m_{2}<m_{3}$\cite{ibarra} from neutrino oscillation
experiments:
\begin{equation}
\mid \varepsilon_{1}\mid \leq \varepsilon_{1}^{max} = \frac{3}{16
\pi }\frac{M_{1}}{v^{2}}\frac{\Delta m_{atm}^{2}+\Delta
m_{sol}^{2}}{m_{3}} \simeq 5.27 \times
10^{-18}\frac{(M_{1}/GeV)}{(m_{3}/eV)},
\end{equation}
where we have used $\Delta m_{atm}^{2}=m_{3}^{2}-m_{2}^{2}=2.6
\times 10^{-3}\textrm{eV}^{2}$ for atmospheric neutrinos\cite{ahn}
and $\Delta m_{sol}^{2}=m_{2}^{2}-m_{1}^{2}=7.1 \times
10^{-5}\textrm{eV}^{2}$ for solar neutrinos\cite{ahmad}. We also
calculate $T_{F}$ for $1 \ll K \lesssim 10^{6}$\cite{kolb}
\begin{equation}
T_{F} \simeq \frac{M_{1}}{4.2(\ln K)^{0.6}}
\end{equation}
with
\begin{equation}
K \equiv \frac{\Gamma _{N_{1}}}{H(T)} | _{T=M_{1}}.
\end{equation}
Here $\Gamma _{N_{1}}$ is the decay width of $N_{1}$. Using
\begin{equation}
\Gamma _{N_{1}}=\frac{1}{8 \pi}(h^{ \dagger} h)_{11}M_{1},
\end{equation}
and Eq.(12), we get
\begin{equation}
K=\frac{\tilde{m}_{1}}{m_{\ast}}
\end{equation}
with
\begin{equation}
\tilde{m}_{1}=\frac{(h^{ \dagger} h)_{11}v^{2}}{M_{1}},
\end{equation}
\begin{equation}
m_{\ast}=\frac{16 \pi ^{5/2}g_{\ast}^{1/2}}{3
\sqrt{5}}\frac{v^{2}}{M_{Pl}} \simeq 1.07 \times 10^{-3}eV.
\end{equation}
Since $\tilde{m}_{1} \geq m_{1}$\cite{bari}, we replace
$\tilde{m}_{1}$ by $m_{1}$ in Eq.(19) and then give a lower limit
for $K$
\begin{equation}
K \geq K_{min} \equiv \frac{m_{1}}{m_{\ast}},
\end{equation}
accordingly $T_{F}$ has an upper bound
\begin{equation}
T_{F} \leq T_{F}^{max} \simeq \frac{M_{1}}{4.2(\ln
K_{min})^{0.6}}.
\end{equation}

Assuming that the decay of the lightest heavy Majorana neutrinos,
which are released from the cosmic string loops, can produce a
baryon asymmetry consistent with the CMB observations (1), we obtain
the following restriction
\begin{equation}
\eta _{B}(\varepsilon_{1}^{max},T_{F}^{max})=\frac{28}{79} \times
7.04 \times \varepsilon_{1}^{max} \int ^{T_{F}^{max}} _{T_{EW}}
\frac{1}{s} [\frac{dn_{N_{1}}}{dt}(T)]\frac{1}{TH(T)}dT \geq \eta
_{B}^{CMB}.
\end{equation}
Then the $U(1)_{B-L}$ gauge symmetry-breaking scale can be
estimated by using the above equation in the SD process and QD
process respectively. In the following calculations, we will take
$x=0.5$, $\Gamma =100$, $C=1$, $\mu =\eta ^{2}_{B-L}$, and
$M_{1}=g_{1}\eta _{B-L}$, where the Yukawa coupling $g_{1}
\lesssim 1$ is natural for $M_{1} \ll M_{2},M_{3}$.

In the SD case, using Eq.(6), (11), (14), (15) and (23), we obtain
\begin{displaymath}
\eta _{B}^{SD}(\varepsilon_{1}^{max},T_{F}^{max})=\frac{28}{79}
\times 7.04 \times \varepsilon_{1}^{max} \int ^{T_{F}^{max}}
_{T_{EW}} \frac{1}{s}
[\frac{dn_{N_{1}}^{SD}}{dt}(T)]\frac{1}{TH(T)}dT
\end{displaymath}
\begin{displaymath}
\simeq \frac{28}{79} \times 7.04 \times \varepsilon_{1}^{max}
\times 5.33 \times 10^{-18}N_{N_{1}}^{SD}f_{SD}
\frac{(T_{F}^{max}/GeV)^{3}-(T_{EW}/GeV)^{3}}{(\eta_{B-L}^{SD}/GeV)^{2}}
\end{displaymath}
\begin{equation}
\simeq 9.44 \times 10^{-37}g_{1}^{4}N_{N_{1}}^{SD}f_{SD}
\frac{(\eta_{B-L}^{SD}/GeV)^{2}}{(m_{3}/eV)[\ln
(m_{1}/m_{\ast})]^{1.8}}.
\end{equation}
In the last step we have neglected the effect of $T_{EW} \simeq
100 \textrm{GeV}$ since the dominant contribution to the integral
comes from $T_{F} \gg T_{EW}$. Considering the constraint (24)
from the CMB, we can get a lower limit for $\eta_{B-L}$
\begin{equation}
\eta_{B-L} \gtrsim 2.39 \times 10^{13}
\frac{1}{g_{1}^{2}(N_{N_{1}}^{SD})^{1/2}f_{SD}^{1/2}}(m_{3}/eV)^{1/2}[\ln
(m_{1}/m_{\ast})]^{0.9} GeV
\end{equation}
where the $3 \sigma $ lower limit for $(\eta ^{CMB}_{B})_{low}=5.4
\times 10^{-10}$ has been adopted. Using the relations
\begin{equation}
m_{1}^{2}=\frac{1}{3}( \bar{m}^{2}-\Delta m^{2}_{atm}-2 \Delta
m^{2}_{sol}),
\end{equation}
\begin{equation}
m_{3}^{2}=\frac{1}{3}( \bar{m}^{2}+2 \Delta m^{2}_{atm}- \Delta
m^{2}_{sol}),
\end{equation}
and fixing $ N_{N_{1}}^{SD}$ and $f_{SD}$, we can get the lower
limits for $\eta_{B-L}$ with $\bar{m}$.

In Fig.1 we show $\eta_{B-L}$ as a function of $\bar{m}$ for
$0.06\textrm{eV} \leq \bar{m} \leq 1.0\textrm{eV}$ (corresponding
to $10 \lesssim k \lesssim 300$) with $g_{1}=1.0,0.1,0.01$ by
taking $N_{N_{1}}^{SD}=1$, $f_{SD}=1$. We find that in order to
satisfy the CMB constraint $\eta_{B-L} \lesssim 1.0 \times
10^{16}\textrm{GeV}$, the Yukawa coupling $g_{1} \gtrsim 0.1$. In
the case of degenerate neutrino scenario with $g_{1} = 0.1$, we
obtain $\eta_{B-L}\gtrsim 3.3 \times 10^{15} \textrm{GeV}$ for
$\bar{m}=0.2 \textrm{eV}$ (corresponding to the upper bound for
the successful leptogenesis\cite{bari}), and $\eta_{B-L}\gtrsim
5.3 \times 10^{15} \textrm{GeV}$ for $\bar{m}=0.4 \textrm{eV}$ or
$\eta_{B-L}\gtrsim 9.5 \times 10^{15} \textrm{GeV}$ for
$\bar{m}=1.0\textrm{eV}$ (corresponding to the upper bounds from
the cosmological data\cite{spergel,tegmark}).

\begin{figure}
\includegraphics[scale=1.4]{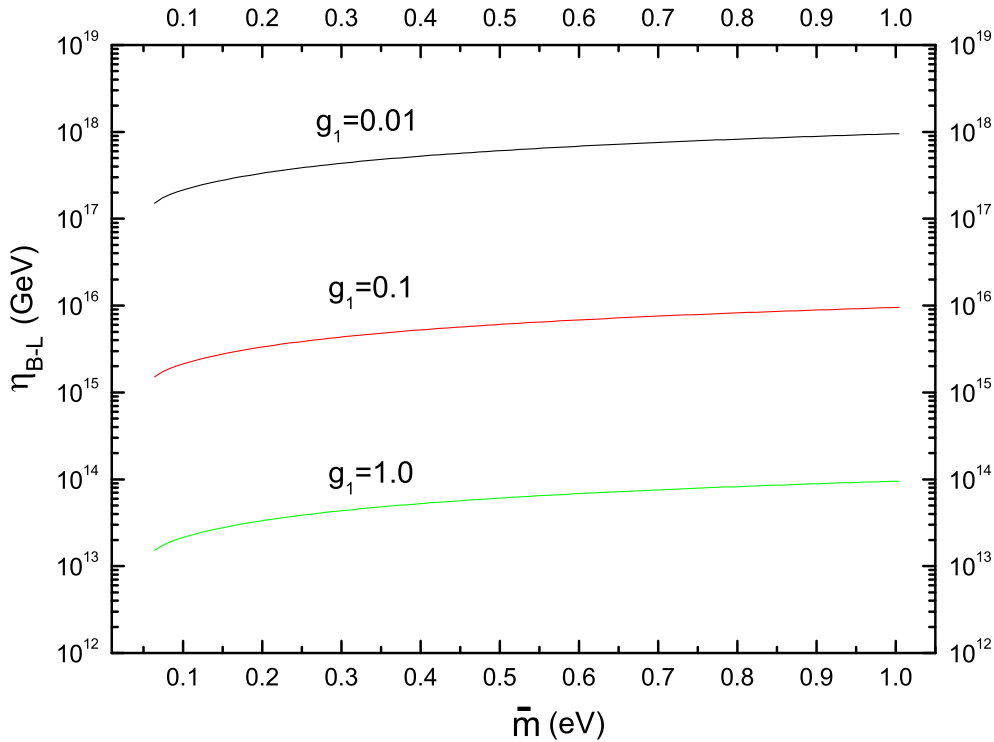}
\caption{The evolution of the lower limits for $\eta_{B-L}$ with
$0.064 \textrm{eV} \leq \bar{m} \leq 1.0 \textrm{eV}$ in the SD
process for $ N_{N_{1}}^{SD}=1$, $f_{SD}=1$.}
\end{figure}

Replacing $\frac{dn_{N_{1}}}{dt}$ in Eq.(14) by
$\frac{dn_{N_{1}}^{QD}}{dt}$ given by Eq.(7), considering the
additional constraint on $f_{QD}$ from Eq.(5) and repeating the same
steps in the SD case above, we can also obtain the baryon asymmetry
and the lower limit for $\eta_{B-L}$ in the QD process
\begin{equation}
\eta _{B}^{QD}(\varepsilon_{1}^{max},T_{F}^{max}) \simeq 4.05
\times 10^{-27}g_{1}^{2}N_{N_{1}}^{QD} \frac{1}{(m_{3}/eV)[\ln
(m_{1}/m_{\ast}]^{0.6}} (\eta _{B-L}/GeV),
\end{equation}
\begin{equation}
\eta_{B-L} \gtrsim 1.33 \times
10^{17}\frac{1}{g_{1}^{2}N_{N_{1}}^{QD}} (m_{3}/eV)[ \ln
(m_{1}/m_{ \ast })]^{0.6}GeV.
\end{equation}

In Fig.2 we plot $\eta_{B-L}$ as a function of $\bar{m}$ for
$g_{1}=1.0,0.1,0.01$ by taking the parameters as $
N_{N_{1}}^{QD}=1$. We find that the lower limits for $\eta_{B-L}$
with $0.06 \textrm{eV }\leq \bar{m} \leq 1.0 \textrm{eV}$ in the
QD process are much higher than the values in the SD process, and
higher than the upper bound $ 1.0 \times 10^{16} \textrm{GeV}$.
Furthermore by taking $\eta _{B-L}=1.0 \times 10^{16}
\textrm{GeV}$ and $ N_{N_{1}}^{QD}=1$ in Eq.(29), we plot the
evolution of $\eta _{B}^{QD}$ as a function of $\bar{m}$ in Fig.3.
We can see that the contribution from the $B-L$ cosmic string
loops to the BAU is small enough to be neglected with the above
neutrino masses for the SD case.

\begin{figure}
\includegraphics[scale=1.4]{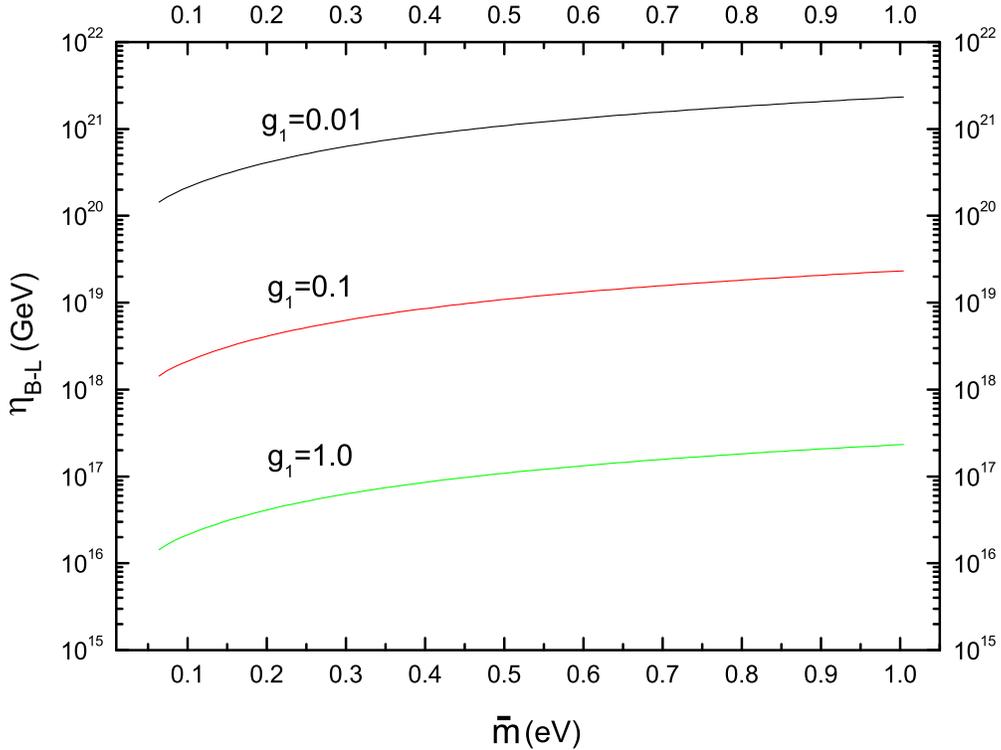}
\caption{The evolution of the lower limits for $\eta_{B-L}$ with
$0.064 \textrm{eV} \leq \bar{m} \leq 1.0 \textrm{eV}$ in the QD
process for $ N_{N_{1}}^{QD}=1$.}
\end{figure}

\begin{figure}
\includegraphics[scale=1.4]{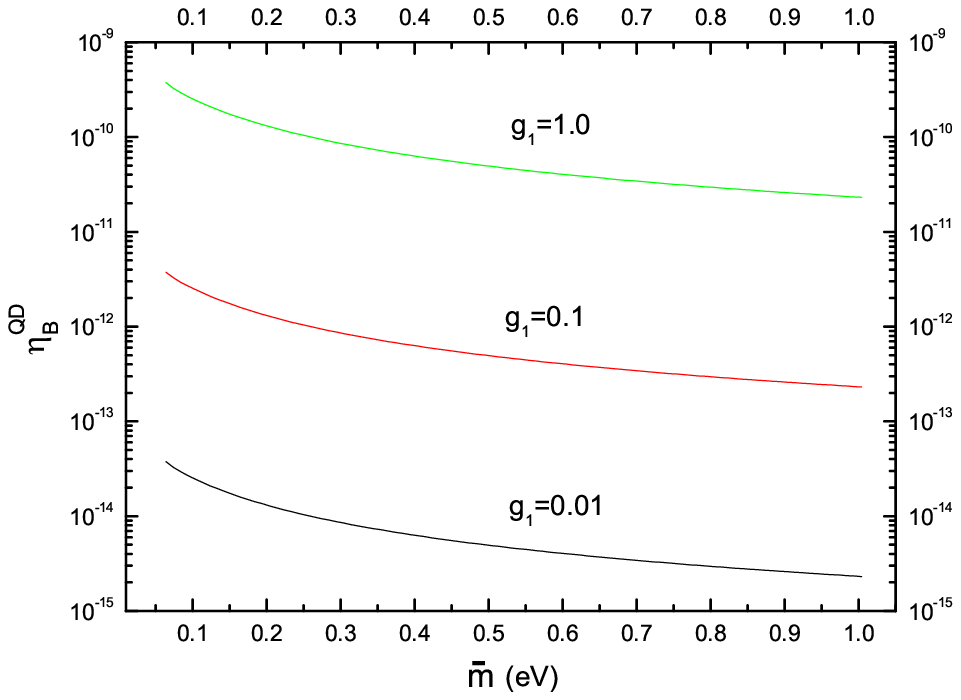}
\caption{The evolution of $\eta _{B}^{QD}$ with $0.064 \textrm{eV}
\leq \bar{m} \leq 1.0 \textrm{eV}$ in the QD process for $
N_{N_{1}}^{QD}=1$ and $\eta_{B-L}=1.0 \times 10^{16}
\textrm{GeV}$.}
\end{figure}

In summary, we study such leptogenesis scenario: the lightest
heavy Majorana neutrinos are released from the $B-L$ cosmic string
loops, and their decay can produce a baryon asymmetry consistent
with the CMB observations. Considering the fact that both the
neutrinoless double beta decay experiment and the cosmological
data show a preference for the degenerate neutrinos, we give the
lower limits for the $U(1)_{B-L}$ symmetry-breaking scale
$\eta_{B-L}$ with $0.06 \textrm{eV} \leq \bar{m} \leq 1.0
\textrm{eV}$, where the full possible cases of degenerate
neutrinos are included. Especially we plot the lower limits for
$\eta_{B-L}$ with $ \bar{m} $ in the SD process and QD process
respectively.

In the SD process, we find that the Yukawa coupling $g_{1}$ should
be $ \gtrsim 0.1$ due to the CMB constraint $\eta_{B-L} \lesssim
1.0 \times 10^{16} \textrm{GeV}$. In the case of degenerate
neutrino scenario, we obtain $\eta_{B-L}\gtrsim 3.3 \times 10^{15}
\textrm{GeV}$, $5.3 \times 10^{15} \textrm{GeV}$ and $9.5 \times
10^{15} \textrm{GeV}$ for the degenerate neutrino masses
$\bar{m}=0.2 \textrm{eV}$, $0.4 \textrm{eV}$ and $1.0
\textrm{eV}$, respectively. And we also find that there is very
small contributions from the $B-L$ cosmic strings to the BAU in
the QD process.

{ \bf Acknowledgment:} We thank Xiao-Jun Bi, Bo Feng, Zhi-Hai Lin
and especially Xinmin Zhang for discussions. This work is
supported partly by the National Natural Science Foundation of
China under the Grant No. 90303004.

\end{document}